\documentstyle[12pt,epsf]{article}

\begin{document}

\begin{flushright}
SLAC--PUB--8151\\
May 1999\\
T,Noyes\\ 
\end{flushright}

\bigskip\bigskip
\begin{center}
{\bf\large
SCIENCE and PARANORMAL PHENOMENA
\footnote{\baselineskip=12pt
Work supported by Department of Energy contract DE--AC03--76SF00515.}
\footnote{\baselineskip=12pt
Conference Proceedings for ANPA 20 will be available from
ANPA c/o Prof.C.W.Kilmister, Red Tiles Cottage, High Street, Barcombe,
Lewes, BN8 5DH, United Kingdom.}}

\bigskip

H. Pierre Noyes\\
Stanford Linear Accelerator Center\\
Stanford University, Stanford, CA 94309\\
\end{center}
\vfill
\begin{abstract}
In order to ground my approach to the study of paranormal phenomena,
I first explain my operational approach to physics, and to the
``historical'' sciences of cosmic, biological,
human, social and political evolution. I then indicate why I believe
that ``paranormal phenomena'' might --- but need not ---
fit into this framework. I endorse the need for a new theoretical
framework for the investigation of this field presented by Etter
and Shoup at this meeting. I close with a short discussion
of Ted Bastin's contention that 
paranormal phenomena should be {\it defined} as contradicting
physics.
\end{abstract}
\vfill           
\begin{center}
Contribution to the discussion of science and the paranormal\\
sponsored by the Epiphany Philosophers as an annex to the\\ 
$20^{th}$ annual international meeting of the\\
{\bf ALTERNATIVE NATURAL PHILOSOPHY ASSOCIATION}\\
Wesley House, Cambridge, England, September 7-8, 1998\\
\end{center}
\vfill
\begin{center}
To be published by ANPA, May 1999, as\\
``Paranormal Aspects: Appendix to the Proceedings of ANPA 20''\\
Available from ANPA, c/o Dr. K. Bowden, Theoretical Physics Research Unit\\
Birkbeck College, Mallet St. London WC1E 7HX\\
\end{center}

\vfill

\newpage

\section{INTRODUCTION --- NORMAL SCIENCE}

There was difficulty during these discussions reaching any consensus on
what was meant by ``paranormal''. In the end we did not try. I suspect
that part of the problem was that our diverse group does not agree on
what is ``normal science'', making a sharp, contrasting definition of
``paranormal'' phenomena impossible for us in the first place.  I have
therefore decided that, before I can explain to you how I try to 
think about paranormal phenomena, I must first explain how I think about ordinary science.

I am a physicist. For me, as for many others, physics is an empirical
science based on quantitative measurements mutually agreed on by a
community of practitioners of physics. That such a community exists,
but has come into existence only since the ``scientific revolution'' of
the seventeenth century, I take to be an established historical fact.
In this sense, I take agreed upon laboratory protocol and practice to be primary
and the mathematical language and other technical terms used in
describing how, up to a point, agreement between members of the
community is achieved to be secondary. Both evolve over time, and bring
in other communities, as is well illustrated by Peter Galison's
incisive examination of the objects on the laboratory floor which
constitute the material culture of particle physics in this century
\cite{Galison97}.
 
What concerns me here is not so much particle physics per se, but
how its conclusions are extended to provide a framework with which to
describe the past. Since I have presented at this meeting
the cosmological framework that comes out of Program Universe
and its connection to bit-string physics\cite{Noyes98}, I will be brief. 
The basic assumptions are: a) the Galilean assumption that processes we
observe occurring here and now will --- until we have evidence to the
contrary --- occur a similar way under similar circumstances elsewhere
in the cosmos; b) the assumption that (except under special circumstances
described by the General Theory of Relativity) light travels at the
limiting velocity $c$ if unimpeded by matter; c) on a large enough
scale (which has now been achieved, thanks to the Hubble Space
Telescope) the universe at any epoch is homogeneous and isotropic,
leading to the Friedman-Robertson-Walker metric for the macroscopic
framework into which we fit our observations. Extrapolating back to 13
billion years from the present (thanks to a number of
recent developments \cite{Noyes98}) now provides a consistent description
of the evolution of the cosmos within our event horizon, with a number
of detailed cross-checks. 

This sounds like a departure from my
commitment to an operational stance about space and time. So I
emphasize that this picture only refers to physical phenomena we can
measure and/or observe here and now. I remain skeptical, even doubtful, as to
whether these successes establish the ``reality'' of space and time in 
any deep sense. Clearly, as with ``common sense'' space and time, they
form a useful descriptive framework, if we do not commit the error of
casting it in concrete. I consider it a real triumph of the ANPA
program that we can arrive at this framework from the combinatorial
hierarchy construction via program universe \cite{Noyes98} or any similar algorithm
{\it without} postulating any {\it a priori} space time.

Granted this background, the older story\cite{Noyes74} of the origin of the solar
system, of biomolecular chirality and biopoesis
\cite{Rubensteinetal83,Bonner92}, and of terrestrial
biological evolution falls into its appropriate niche. Recent work,
which I will not bother to cite, has enormously deepened and enriched
this description and (for me, at least) strengthened my conviction
that no major lacunae remain. I stress that the ``here and now''
sciences --- physics, chemistry, biology, ... --- are a necessary background
for understanding the historical sciences in the broad sense: cosmology,
stellar and solar system evolution, terrestrial biological evolution,
evolution of human intelligence and language, social evolution,
political evolution. As we proceed up the chain from
physics to politics, the scientific disciplines  
become more and more contingent on unique,
local events whose prevalence in the rest of the cosmos we
can currently only guess at. However, the recent discovery of many 
extra-solar planetary systems in our immediate neighborhood makes it
possible that, in the not too distant future, some of these guesses
about exobiology may be replaced by hard fact. We may also be on the
threshold of understanding the co-evolution of language and the brain
in the human species if Deacon\cite{Deacon97}, among others, is to be believed.     
  
\section{WHAT ABOUT PARANORMAL PHENOMENA?}  

Much recent work on ``paranormal phenomena'' has amounted to getting
large statistical samples with small deviations from ``chance'' which
are unexplained. Much of this work has considerably higher
methodological standards than most scientific work. However, for those
familiar with experimental physics (and presumably in many other fields
as well) this will never be convincing.  We are all too familiar with
unexplained effects that cannot be attributed to ``chance''. For
us these are examples of systematic error, and if they cannot be
brought under control, simply characterize a bad experiment.
One has to {\it understand} the sources of systematic error, show
that they vary in a systematic way with changes in experimental
conditions, and do one's best to bring them down {\it below} the effects
of statistical error. For this, of course, one needs a theory, not only
of the phenomenon being investigated, but {\it also} a theory of what
is (or is likely to be) interfering with the measurement. I do not see
how this situation can be achieved in investigations of paranormal
phenomena without much more theoretical work
using a framework that allows for the testing of hypothesis
{\it and their rejection}.  In this I agree with what Etter and Shoup
have already said at this meeting, and in this discussion. 
But I would
go further and say that one needs not only a quantitative theory for the phenomena
themselves, but {\it also} a theory for sources of systematic error
in a form which can also be tested. 

The impetus for research into
paranormal phenomena has not come, and does not now come, from small,
inexplicable effects. Judging by material presented in this discussion,
and from my own contacts with scientists interested in the subject, 
I assert that this interest usually arises from personal experience.  
I have never had any ``paranormal'' experience. But people I respect,
including some at this meeting, tell me they have. So I take the
possibility that some people have this capacity seriously. I also
do not get much out of listening to music. But I have plenty of
evidence that many people do. In both respects I am not unusual.

I start with an incident I heard of three decades ago, which was told
to me by an anthropologist\cite{Noyes82}. In brief, while working one day
in the Pacific Northwest with a shaman he had known for several months,
the shaman asked suddenly if the anthropologist would like to
know what the anthropologist's friend in Chicago was doing just then.
Of course he said yes. Equipped with the shaman's response,
the anthropologist documented it, wrote to his friend in Chicago and
got a statement of what he was doing at that time. The correspondence
between the shaman's report and the friend's statement was so close
that the anthropologist, twenty years later, was still afraid to
publish for fear it would damage his professional reputation.

I didn't know what to do with this story at the time. However a year
or so later I proved that when a system with two quantum mechanical
particles interacting via short range forces is augmented by a third
particle with similar interactions, the behavior of the pair changes
no matter how far away the third particle is. I called this example
of the extreme non-locality of quantum mechanics the {\it eternal
triangle effect}, and compared it analogically with the above instance
and other behavioral examples. The analysis I subsequently
published\cite{Noyes82} provides a good starting point for discussing my
current position. I quote:

\begin{quotation}

It is not necessary for you to believe the story in order to ask the
question, as I do, of how such a remarkable `communication' might
occur. After much rumination on the event, and after the discovery of
the eternal triangle effect and its behavioral analog, I have come to a
tentative model, or rather explanatory framework. Since the
anthropologist and the shaman had reached a mutual level of confidence
and trust, they could to a certain extent `share each other's
thoughts' --- [a] phenomenon known to all of us, and not necessarily
involving any paranormal phenomena
\footnote{It is relevant here that the anthropologist was one of the
founders of kinesics; he once told me that given only a minute or so of
the start of a filmed psychotherapeutic session, he could predict
what would happen during the rest of the hour.}.
Further, the anthropologist knew his distant friend well, and might by
[a] similar process anticipate (unconsciously) what his friend would be
doing at the time. We know of many instances when such unconscious
deductions come to us in dreams --- sometimes accurate and sometimes
not. For the shaman to `pick up' this knowledge or conjecture from the
anthropologist need involve only the types of `non-verbal communication'
discussed in this volume. and which, though often difficult to
understand, model, or demonstrate, are again familiar aspects of human
behavior. Granted only the postulate that a human mind makes makes many
accurate deductions about present [and future] happenings
from past experience --- which would shock no psychoanalyst --- the
whole incident can be fitted into the framework of explanatory models
that, separately, are often accepted.

It is interesting to speculate on whether many phenomena which are
called `paranormal' might not fit into such an explanatory framework.
The `framework' does not really explain anything, of course. To account
for an unexplained occurrence by saying that the human mind can make,
unconsciously, very accurate deductions about what will occur
(`precognition'), what another person is thinking (`telepathy'),
or how an unstable system will behave (predictive `telekinesis')
is only to replace one problem with another --- namely how to explain
this extraordinary computational ability. But it does have the aspect
of explaining a fact that is troublesome in `paranormal research',
namely that the ability is not 100\% and closely tied to the
emotional state of the individual\footnote{In the light of our
discussion of systematic error above, it occurs to me that
`emotional state' of both subject and experimenter is one
factor that cries out for quantitative assessment and investigation
in this field --- perhaps an impossible task?}. This is what we would
expect, from psychoanalytic theory, of a process deeply buried in the
unconscious. Coming back to the theme of this volume, such unconscious
processes clearly can have an important bearing on non-verbal
communication of more conventional sorts, and it is perhaps reassuring
that the underlying physics warns us we should include them in our
thinking about how such communications work.

My intention in this essay is not to say that quantum mechanics
`explains' paranormal phenomena by some such route. What I do claim is
that quantum mechanics, in the simplest case where the 
phenomena can occur (the three particle problem with finite range
interactions), {\it does} require both an extreme nonlocality of
description when forced into an `instantaneous' or `static' form,
and the inclusion (in principle)  of {\it all} past events in the
discussion of the current situation. I hope that this fact can provide
an `explanatory framework' within which it is easier to contemplate
correlations between events so distant in space and time from each other
as to make models drawn from classical physics seem inadequate or
implausible.

\end{quotation}
   
My first criterion for the establishment of a scientific study of
paranormal phenomena is that it be capable of convincing skeptics
like me that meaningful experimental investigation is possible in the
first place. If the investigations are statistical, it is all too easy
to dismiss their results as due to unexplained systematic error.
If they are anecdotal, it is all to easy to fall back, as I have done
in the analysis just quoted, on some form of unexplained
``unconscious'' effect that falls more properly in the domain studied
by psychiatrists than in a new discipline. 

I am afraid that all too
many ``scientists'' are uncomfortable living in a world in which
most of the important things in life are unexplained, and grasp
at facile explanations or rejections. For me the true scientist 
lives with uncertainty as his constant companion, and never expects that
situation to change. But that does {\it not} mean that new facts and
methods are to be avoided; rather, they should be eagerly pursued.
I look at one new possibility in the next section.

\section{A NEW METHODOLOGY?}

We have already heard from Etter and Shoup about a new approach to the
study of paranormal phenomena based on new theoretical insights that 
have come out of  Etter's work on the foundations of quantum mechanics.
Since the up to date material will not be available for a while in
written form, I refer you to an older paper of Tom's, which is now
available on the web\cite{Etter&Noyes98}. What Tom does is to show that
the core laws of quantum mechanics (Born's probability rule that
gives probabilities as the squares of ``amplitudes'', and the unitary
evolution of the quantum state called Schroedinger's equation) are
simply a piece of mathematics which has no physics in it. This
allows him to formalize the Markov chains (which are irreversible)
with either past or future boundary conditions, and use the same
framework to describe the time-reversible Schroedinger evolution.
Thus classical (statistical) systems peacefully coexist with quantum
systems, as they must in quantum measurement theory. Hopefully,
his discussion of quantum measurement theory will make this subject
less paradoxical for some who have trouble with it. Although his
approach provides a new way of looking at quantum mechanics, at this
stage no new predictions are made.

What makes Etter's analysis exciting from the point of view of this
paper is that in addition to quantum mechanics, the formalism allows
a clean description of phenomena, such as ``future causation'',
which appear to occur in many reports of paranormal phenomena.
But this descriptive framework, being general, is not tied to
Planck's constant. Thus it provides for the possibility of macroscopic
acausality which, as I indicated in the last section, is analogically
suggested by quantum mechanics, but without giving a clue as to how
to make a systematic theory for it. 

Even having a theory is useless, except as an aid to imagination, 
until a way is found to fit experimental results into the theoretical
framework. The payoff is when experimental results thus formulated
lead to a reliable technology which can join the everyday world of
fact. I must confess that I am skeptical whether this can be done
for paranormal phenomena,
but I enthusiastically support Etter and Shoup's efforts to take this
step.  

\section{CONCLUSION}

I conclude by turning to Ted Bastin's proposed definition
of the paranormal. He started with the
proposition that paranormal phenomena show no dependence on space and
time. He then coupled this to his further assumption that current
physics {\it begins} with space and time. These two propositions
in conjunction make a clash with normal science inevitable.

This need not be the case. I am not the only contemporary physicist 
who feels the need
to {\it construct} space and time as part of the foundations of
physics. Since I think of myself as doing ``normal science'', 
or posibly as encouraging a paradigm shift which will turn out 
to be acceptable by normal scientists, I cannot
accept the second half  of Ted's position. I quite
agree with Ted that {\it many} scientists do start by uncritically
accepting either the continuum space-time of physics or the cruder
space-time of ``common sense'' as the given theatre in which the dramas
they study take place. But I do not go along with them.
In fact many people accept the fact that demonstrated macroscopic
quantum phenomena such as supraluminal correlation without supraluminal
signalling over distances of 20 kilometers, 
and the teleportation of photons (destruction at one
position and recreation of the same photon at a separate space-time
location) show that ``space-time'' is more complicated than the 
Maxwellian picture allows for. Similar remarks could be made about
black holes and modern cosmology.

Thus, for me, it comes down to whether the best strategy for getting on
with the job of obtaining a better understanding of ``paranormal
phenomena'' is to follow a course that inevitably leads to
confrontation, or to find a way to expand ``normal science'' so that
it can include such phenomena. Obviously, from what I have said in this
paper, I currently favor the latter course. But I am ready to be convinced
that this is a mistake. 

I end by giving my heartfelt thanks to the Epiphany Philosophers for
making possible these two days of  very interesting discussion.

\end{document}